# A global summary of seafloor topography influenced by turbulent water mixing

by Hans van Haren and Henk de Haas

NIOZ Royal Netherlands Institute for Sea Research, P.O. Box 59, 1790 AB  Den Burg, the Netherlands.
e-mail corresponding author: hans.van.haren@nioz.nl

ABSTRACT

Turbulent water motions are important for the exchange of momentum, heat, nutrients, and suspended matter including sediments in the deep-sea that is generally stably stratified in density. To maintain ocean-density stratification, an irreversible diapycnal turbulent transport is needed. The geological shape and texture of marine topography is important for water mixing as most of deep-sea turbulence is generated via breaking internal waves at sloping seafloors. For example, slopes of semidiurnal internal tidal characteristics can 'critically' match the mean seafloor slope. In this paper, the concept of critical slopes are revisited from a global internal wave-turbulence viewpoint seafloor topography- and using moored high-resolution temperature sensor data. Observations suggest that turbulence generation via internal wave breaking at 5±1.5% of all seafloors is sufficient to maintain ocean-density stratification. However most, >90%, turbulence contribution is found at supercritical, rather than the more limited critical, slopes measured at 1′-scales that cover about 50% of seafloors at water depths < 2000 m. Internal tides (~60%) dominate over near-inertial waves (~40%), which is confirmed from comparison of NE-Atlantic data with East-Mediterranean data (no tides). Seafloor-elevation spectra show a wavenumber (k) fall-off rate of $k^{-3}$, which is steeper than previously found. The fall-off rate is even steeper, resulting in less elevation-variance, in a one-order-of-magnitude bandwidth around $k_T=0.5$ cycle-per-km. The corresponding length is equivalent to the internal tidal excursion. The reduction in seafloor-elevation variance seems associated with -erosion by internal wave breaking. Potential robustness of the seafloor-internal wave interaction is discussed.

1. Introduction

The present ocean exists for millions of years, and so seemingly do its seafloor shape and its water flows including properties like the stable vertical stratification in density from the heating by the sun. Because sedimentation and erosion of suspended matter in the ocean are subject to water flows that partially depend on interaction with the seafloor, one may question the stability or variability of seafloor-shape and water flow properties. Compared with geological time scales of variation (sedimentary: years; rocks: centuries/millennia), water flows in the deep ocean are fast fluctuating (meso-scale eddies: weeks/months; tides: hours/days). However, that does not preclude potential interactions between water flows and topographic bed-forms, with local results such as sand ripples and sediment waves (e.g., Trincardi and Normark, 1988; Puig et al., 2007).

As seafloor-erosion by resuspension is mainly driven by intense water turbulence, and -deposition by weak turbulence, seafloor shaping such as contourite morphology (e.g., Rebesco et al. 2014; Chen et al., 2022) depends on dominant ocean-turbulence generation processes. Following, e.g., Wunsch (1970), Eriksen (1982), Thorpe (1987), Klymak and Moum (2003), Hosegood et al. (2004) and Sarkar and Scotti (2017), ocean turbulence above sloping seafloors is predominantly generated via the breaking of internal water waves. Such waves are mainly supported by the stable vertical density stratification. Their effects on shaping ocean's seafloor morphology cannot be underestimated (Rebesco et al., 2014).

More in general, balancing feed-back interaction leading to a quasi-stable equilibrium is expected between the slopes of seafloor topography and long-lasting ubiquitous water-flows, e.g., generated by internal waves that are notably driven by oscillating tides and the waves' density-stratification support. In their pioneering works, Bell (1975a,b) from an internal wave generation perspective, Cacchione and Wunsch (1974) and Eriksen (1982; 1985) from an internal wave breaking perspective, and Cacchione and Southard (1974) from a geological formation perspective, suggested that a 'critical' match exists between the mean angles of deep-sea topography slopes and those of internal wave 'characteristics'.

Because internal water waves are essentially three-dimensional (3D) phenomena, which distinguishes them from 2D surface waves, their energy propagates along characteristics, i.e. paths that slope to the horizontal as a function of wave frequency and stratification. Internal waves can reflect off a seafloor slope, but internal wave energy is thought to build-up when the two slopes are identical, yielding propagation parallel to a critical seafloor slope (Cacchione and Wunsch, 1974). When the seafloor slope is larger than the internal wave slope it is supercritical, when it is smaller than the internal wave slope it is subcritical (for that particular wave frequency under given stratification conditions).

While Bell (1975a,b) considered seafloor elevation statistics on internal wave generation, the works by Cacchione and co-authors considered seafloor shaping by internal wave motions. It was found that the average seafloor slope amounts 3±1°, which matches the characteristics slope of energy propagation direction of internal waves at semidiurnal tidal frequency for stratification from around 1700 m below the sea surface, which is about half the average water depth.

While the above finding is remarkable and has led to numerous investigations on critical reflection of internal waves, it is challenged from various perspectives. This is because: First, exactly matching 'critical' slopes for dominant internal tides are few and far between and cannot persist in space and time because the ocean stratification varies in space and time. Second, internal waves mostly break above sloping topography, very little in the ocean interior (Polzin et al., 1997). Third, most ocean seafloors, i.e. 75% by area and 90% by volume (Costello et al., 2010; Costello et al., 2015), occur between 3000 and 6000 m water depth where density stratification is weak. Fourth, about 40% of internal wave energy is estimated at near-inertial frequencies (e.g., Wunsch and Ferrari, 2004). Near-inertial waves are generated as transients following geostrophic adjustment on the rotating Earth, e.g., after passing atmospheric disturbances or frontal collapses (LeBlond and Mysak, 1978). Their occurrence shows a strong variability with location, e.g., demonstrating large near-surface generation in western boundary flows and much weaker generation in eastern ocean-basins (Watanabe and Hibiya, 2002; Alford, 2003). As near-inertial waves have a quasi-horizontal component, virtually all seafloor slopes are steeper: They are super-critical (for near-inertial waves). These observations demand further investigation in the seafloor/internal wave-turbulence interaction, also considering the potential impact on climate variability and the contribution of oceans in distributing heat therein.

Detailed ocean observations indicate that most intense turbulence and sediment resuspension over sloping seafloors is not generated by frictional (shear-)flows (as suggested by, e.g., Cacchione et al., 2002), but by nonlinearly deformed internal waves breaking (e.g., Klymak and Moum, 2003; Hosegood et al., 2004). Such nonlinear internal wave breaking predominantly generates buoyancy-driven convection-turbulence. Growing evidence suggests that most breaking occurs at slopes that are just supercritical (van Haren et al., 2015; Winters, 2015; Sarkar and Scotti, 2017).

Any interaction between seafloor shape&texture and water-flow turbulence is expected to vary on geological time scales, because besides deep-sea topography also ocean-interior vertical density stratification exists as long as the oceans do: The ocean is not and has not been a pool of stagnant cold water underneath a thin layer of circulating warm water heated by the sun (Munk and Wunsch, 1998). The question is how stable a balance of such an interaction can be, e.g., between internal waves that are supported by varying stratification and seafloor topography. Is the balance an optimum, or rather a marginal equilibrium, like the marginally stable stratification supporting maximum destabilizing internal shear in shelf seas (van Haren et al., 1999)? Especially given relatively rapid changes, such as Earth-surface heating attributed to mankind, does a balance buffer any modifications: to topography (long time scale), vertical density stratification (medium time scale), and/or to water-flow (short time scale)? Prior to being able to (mathematically) predict any potential tipping point of a balance, the physical processes need to be understood that contribute to a balance. Amongst other deep-sea processes, this involves the physics of internal wave-formation and -breaking into turbulence generation upon interaction with topography and restratification of density.

In this paper, we discuss the (im)possibility of relating seafloor statistics with open-ocean density stratification and internal wave breaking as observed in recent measurements. We revisit concepts of deformation, erosion, and sedimentation, of deep seafloor topography in interaction with internal waves, with vertical density stratification, with turbulence, with ocean's heating/cooling, and consider the (in)stability of these interactions. Instead of Bell's (1975a,b) perspective of internal wave generation, above NE-Pacific abyssal hills, we adopt the perspective of internal wave breaking and turbulence generation through moored high-resolution temperature observations above a wide variety of deep-ocean topography. We also adopt a geological seafloor topography perspective using some detailed multibeam echosounder and global seafloor elevation repository data. An attempt is made to consider the statistical spread of different variables.

*1.1. Some considerations on ocean variability and internal wave-topography interactions*

A large variability over two orders of magnitude exists in stable ocean-density stratification ($\sim N^2$), albeit a gradual, not constant, decrease is observed in buoyancy frequency, a measure of stability of a fluid to vertical displacements, N with increasing depth (e.g., Wüst, 1935; Wunsch, 2015). This gradual decrease with depth results in a corresponding increase in the slope β to the horizontal of characteristics along which internal wave energy propagates. For general ocean stratification this slope is approximately inversely proportional to N (e.g., LeBlond and Mysak, 1978),

$\beta = \sin^{-1}((\omega^2-f^2)^{1/2}/(N^2-f^2)^{1/2})$,  (1)

for freely propagating linear internal waves at frequency ω. Here, f denotes the inertial frequency or Coriolis parameter, the vertical component of planetary vorticity. Internal waves and sub-inertial water-flows deform the stratification locally, thereby making it a function of time and space N = N(x,y,z,t). The consequences of variability for β will be explored from observations in Section 3.

Several global internal wave statistics can be given. The (rms-)mean deep-sea topography slopes of 3±1° (Bell, 1975b; Cacchione et al., 2002) roughly match the mean slope of (linear) semidiurnal internal tide characteristics, provided the latter are computed using in (1) a value of $N \approx 2\times10^{-3}$ s$^{-1}$ for mid-latitude locations. Such N is found approximately around z = -1700 m in the open ocean, which has a mean depth of H = 3900 m outside shelf seas (Wunsch, 2015). In these open-ocean mid-depth waters internal wave breaking and thus turbulence generation are sparse (e.g., Gregg, 1989; Polzin et al., 1997; Kunze, 2017).

Observational evidence suggests that vigorous turbulent mixing by internal wave breaking is only found in a limited height of h = 100-200 m (e.g., Polzin et al., 1997) above the seafloor, coarsely estimated to occur over only 5-15% of all seafloor slopes to maintain the ocean stratification (e.g., van Haren et al., 2015). As for relevant length scales: Although satellite altimetry observations demonstrate low-mode internal tides having wavelengths O(100) km (e.g., Dushaw, 2002; Ray and Zaron, 2016), the excursion length of internal tides is typically O(1) km and which may prove important for turbulence generation and thus sediment erosion of seafloor-texture and -topography.

Especially the smaller length scale O(1) km may fit the spectral analysis of NE-Pacific seafloor elevation, which is found to fall-off with horizontal wavenumber (k) like $k^{-2.5}$ (Bell, 1975a), later corrected to $k^{-2}$ (Bell, 1975b). The latter is interpreted as a random distribution of hills in which the energy of formation is distributed uniformly over all sizes. Internal wave generation is found by Bell (1975a,b) mainly between 0.33 < k < 3 cpkm (short for cycles per kilometer). Here, we add that this roughly matches a spectral band-broadening between 0.15 < k < 1 cpkm, which is visible in the presented data albeit not mentioned by Bell (1975a,b).

Geomorphology influences water-flow and turbulence and these in turn influence sediment erosion and deposition and thus (fine-tuning) the geomorphology. So, if enough time is available (after all, geologists think in terms of millions of years), one will eventually reach an equilibrium. The major geomorphological processes occur on a much larger timescale than the adaptation of water-flows, and any ocean will "always" be in an equilibrium situation because the flows adapt relatively quickly to a slow geological change.

Thus, following variations in ocean internal wave turbulence, seafloor topography will adapt. However, the ocean also interacts with a faster adaptive/varying system: the atmosphere. Temperature changes in water are slower than in air because of the larger heat capacity of the former. Nevertheless, a direct correspondence with changes in vertical density (temperature) stratification is not evident as larger stratification can support more internal waves and thus potentially more turbulent wave breaking that may restore a balance. (It is noted that atmosphere dynamics is not driven by the ocean, except indirectly by modification of moisture content).

The dominant source of ocean internal waves are tides (e.g., Wunsch and Ferrari, 2004). The local seafloor slope $\gamma$, computed over a particular horizontal distance, is supercritical for linear freely propagating semidiurnal lunar $M_2$ internal tidal waves, when $\gamma > \beta_{M2}$, using $\omega_{M2} = 1.405 \times 10^{-4}$ s$^{-1}$ in (1).

A secondary source are near-inertial waves $\omega \approx f$, which generally have a near-horizontal slope of characteristics. Only in very weak stratification $N = O(f)$, some near-inertial slopes may become large enough to distinguish various subcritical slopes for such waves. However, under weakly stratified conditions, terms involving the horizontal Coriolis parameter are no longer negligible, and two distinctly differently sloping characteristics $\mu_{\pm}$ of internal wave energy propagation result (e.g., LeBlond and Mysak, 1978; Gerkema et al., 2008):

$$\mu_{\pm} = (B \pm (B^2 - AC)^{1/2})/A, \qquad (2)$$

in which $A = N^2 - \omega^2 + f_s^2$, $B = ff_s$, $C = f^2 - \omega^2$ and $f_s = f_h \sin\alpha$, $\alpha$ the angle to latitude ($\varphi$). The slopes of $\mu_{\pm}$ indicate directions to the horizontal in a more general way than the single slope (1) obtained under the traditional approximation.

For large $N \gg f$, the slope of the two characteristics in (2) approach each other and their slope approaches $\beta$ in (1). Under conditions $N < 10f$ and latitudinal propagation ($\alpha = \pi/2$), one of the characteristics in (2) becomes quasi-horizontal for which virtually all seafloor slopes are supercritical, and the other becomes more steeply sloping.

The impact of (2) may also not be ignorable for semidiurnal tides in weakly stratified waters around mid-latitudes. With the full non-approximated equations (2) and a stratification of about $N = 8f$, which is typical in waters near the 3900-m mean depth of the ocean seafloor, the slope (1) of $\beta_{M2}$ = 9.1° rather becomes 10.3° and 7.9° for up- and down-going characteristics $\mu_\pm$, respectively (for $|\varphi|$ = 37°).

This spread of about ±13% is a substantial addition to variation in seafloor slope-criticality and becomes larger at weaker stratification (N < 8f) and smaller at stronger stratification (N > 8f). Although the vertical density stratification is generally a monotonic decreasing function with increasing depth, deviations occur, such as in some, e.g. equatorial, areas around 4000 m where larger N is found than above and below, and which is attributed to the transition between deep Arctic waters overlying most dense Antarctic waters (King et al., 2012).

The ubiquitous linear internal waves at various frequencies f ≤ ω ≤ N (for N > f) provide ample options for wave-wave and wave-topography interactions. While some interactions seem too slow, the wave-topography interactions above sufficiently steep topography show strongly nonlinear wave deformation resulting in convection-turbulence when the ratio of particle velocity over phase speed u/c > 1: A fast process. Propagation of such highly nonlinear internal waves is beyond the scope here, but the turbulence dissipation rate of internal wave energy affecting seafloor sediment is not.

Although the ocean and deep-seas are overall turbulent in terms of large bulk Reynolds numbers Re well exceeding Re > $10^4$ and more generally Re = $O(10^6)$, it is a challenge to study the dominant turbulence processes. As the ocean is mainly stably stratified in density, which hampers the vertical size-evolution of fully developed three-dimensional isotropic turbulence, it is expected that in the deep-sea stratification is much weaker resulting in near-neutral conditions of (almost) homogeneous waters in which N ≈ f. Under such conditions turbulent overturns may be slow and large and may govern more the convection-turbulence process, rather than the shear-turbulence process that dominates under well-stratified conditions N >> f and in frictional flows over the seafloor.

*1.2. Internal wave energy dissipation perspective*

According to Wunsch and Ferrari (2004), in follow-up from Munk and Wunsch (1998), the currently best estimate for global internal wave power to be dissipated is 0.8 TW (1 TW =$10^{12}$ W) for internal tides and about 0.5 TW for wind-enforced mainly near-inertial waves. These numbers are determined to within an error of a factor of 2, although this error range is probably smaller for internal tides (since the rather precise determination of energy loss of the Moon-Earth system).

If we distribute this amount of power over the entire global ocean with a surface of $3.6\times10^{14}$ m$^2$ (e.g., Wunsch, 2015), the vertically integrated dissipation rate amounts for internal tides,

$$2.2\times10^{-3} \text{ W m}^{-2} = \int\rho\varepsilon \, dz, \tag{3}$$

and 60% of that value for inertial waves. In (3), $\rho$ = 1026 kg m$^{-3}$ denotes an average density of ocean-water and $\varepsilon$ the kinetic energy dissipation rate.

If we suppose that (3) is distributed over the entire vertical water column, over a mean water height of H = 3900 m (Costello et al., 2010; Wunsch, 2015), a global mean rate to dissipate the internal tidal energy is required of,

$$\varepsilon_H = 6\times10^{-10} \text{ m}^2 \text{ s}^{-3}, \tag{4}$$

and 60% of this value for near-inertial waves. As a result, the entire global-mean turbulence dissipation rate of all internal waves (generated by internal tides and near-inertial waves) is about $10^{-9}$ m$^2$ s$^{-3}$. This is equivalent to the mean value found after evaluation of 30,000 ocean profiles on internal wave turbulence (Kunze, 2017).

Given a mean mixing efficiency of $\Gamma$ = 0.2 (Osborn, 1980; Oakey 1982; Dillon, 1982), and N = $1.5\times10^{-3}$ s$^{-1}$ found around open-ocean z = -1900 m, one arrives at a vertical (actually, diapycnal) turbulent diffusivity of $K_z = \Gamma\varepsilon N^{-2} = 10^{-4}$ m$^2$ s$^{-1}$, for above global internal-wave-induced dissipation rate. This is the canonical $K_z$-value proposed by Munk (1966) and Munk and Wunsch (1998) to maintain the ocean stratification and to drive the meridional overturning circulation.

However, according to measurements using extensive shipborne water column profiling (e.g., Gregg, 1989; Kunze, 2017) and some moored high-resolution temperature sensors (van Haren, 2019) the average open-ocean dissipation rate amounts $4\pm2\times10^{-10}$ m$^2$ s$^{-3}$, which is less than half the required value to maintain the ocean stratification. Locations are thus sought where turbulent mixing is sufficiently strong to cover at least $6\times10^{-10}$ m$^2$ s$^{-3}$ for the insufficient turbulent mixing by sparse internal wave breaking in the open-ocean interior.

It has been suggested that >99% of overall internal wave related $\varepsilon$ is to be found for -2000 < z < -380 m (Kunze, 2017), reasoning that in this depth zone stratification, $N^2 \propto \varepsilon/K_z$, is largest. However, $\varepsilon$ and $K_z$ are not necessarily (un)related, and more complex correspondence has been observed between the three parameters $\varepsilon$, $K_z$ and N, e.g. in Mount Josephine data (van Haren et al., 2015). Above particularly sloping seafloors, turbulence dissipation rate is found to increase with depth (e.g., Polzin et al., 1997; van Haren et al., 2015; Kunze, 2017). The internal wave breaking potency above the abundant seafloor topography led Armi (1979) and Garrett (1990) to propose that one-and-a-half orders of magnitude larger turbulence than found in the ocean-interior would be needed in a layer O(100) m above all seafloors. This suggestion did not include the particulars of dependency internal wave turbulence intensity on stratification, slopes, and wave-nonlinearity.

Internal waves, in particular internal tides, have amplitudes of several tens of meters, which in the vicinity of sloping topography may grow over 50 m, whereby they deform nonlinearly. So, if we suppose a breaking zone of h = 100 m, one needs above all seafloors local turbulence intensity of the value of (4) augmented by a factor of H/h = 3900/100,

$$\varepsilon_h = 2.3\pm0.7\times10^{-8} \text{ m}^2 \text{ s}^{-3}. \tag{5}$$

This value has been observed above a (semidiurnal tidal) critical slope around H = 2500 m of Mount Josephine, NE-Atlantic Ocean (van Haren et al., 2015). But, not all seafloor-slopes show the same level of internal wave breaking, and variations in turbulence dissipation rate by a factor of 100 have been observed between sub- and supercritical slopes over horizontal distances of only O(10) km.

Potential high-turbulence locations are supercritical slopes (Winters, 2015; van Haren et al., 2015; Sarkar and Scotti, 2017) and canyons (van Haren et al., 2022), where moored observations demonstrate one order of magnitude larger tidally averaged values than (5) of,

$$\varepsilon_{ho} = 3.5 \pm 1 \times 10^{-7} \text{ m}^2 \text{ s}^{-3}, \tag{6}$$

due to internal tidal and near-inertial wave breaking across a larger observational height of $h_o$ = 200±50 m above seafloors around H = 1000±200 m. About 60% of turbulence dissipation rate occurs in half an hour during the passage of an upslope propagating bore with 50-m averaged peak intensities of $10^{-5}$ m$^2$ s$^{-3}$ (van Haren and Gostiaux, 2012).

If such turbulence intensity as in (6) occurs in 250/3900 of mean water depth it needs to occur over only 5±1.5% of all slopes to sustain 1.6 times (4). This 5% is still a considerable portion of the ocean's seafloors. If just by internal tides, because virtually all seafloors are supercritical for near-inertial waves, supercritical slopes are required to comprise 3±1% of the slopes, according to (4). In half-shallower waters of H = 1900 m, these percentages of slopes are reached for similar turbulence intensity over h = 125 m, but these shallower water depths present only about 10% of the ocean seafloor area (Costello et al., 2010; Costello et al., 2015). We elaborate in Section 3. It is noted that we require (just) supercritical slopes for intense turbulent internal wave breaking, not critical slopes that are limited over (much) smaller areas and are prone to vary more with space and time than supercritical slopes.

*1.2.1. Variability in linear internal waves*

Considering the limited occurrence of critical slopes, we address the variability in internal (tidal) wave slopes (1), $\beta = \beta(\omega, f, N, t)$. At a fixed mooring location, planetary f (f$^p$) has zero variability, but relative rotational vorticity of up to f$^r$ = ±0.05f$^p$ may be introduced by (sub-)mesoscale eddy activity, so that f should be replaced by 5% variable local effective Coriolis parameter f$^{eff}$ = f$^p$ + f$^r$ (e.g., Kunze, 1985). Likewise, different semidiurnal tidal internal wave frequencies lead to an, e.g. spring-neap, variation of slopes. Because solar frequency S$_2$ differs by 3.5% from lunar M$_2$, $\Delta\beta(\omega)$ varies by about 6% around mean $\beta$. (It is noted that internal $M_2$ cannot propagate freely poleward of $|\varphi| > 74.5°$). Natural variability in density stratification, by a complex of varying flows at internal wave, (sub-)mesoscale, seasonal, and decadal scales leads to variations in mean 100-m-scale N of 5-10% and to local variations of up to 20% in small-scale layers: A lot depends on the particular vertical length scale used in the computation of N, which should last at least a buoyancy and better an inertial period.

Summing up, overall variations of 10% in β are common for linear semidiurnal internal tides. The associated variation of characteristics slope angle of 0.5°, for mean N = 10f, yields a variation O(100) m over a horizontal distance of 10 km. These amounts double when non-traditional effects (2) are considered, and thereby become of the same order of magnitude as the typical internal tidal excursion length. A precise localization of persistent "critical" slopes is therefore not possible.

From a geological perspective one may question how, and in what stable equilibrium the shape of topography exists, as internal tides depend on frequency and latitude, but foremost on underwater vertical density stratification. In addition, (linear) internal tides can deform after interaction with other internal waves such as those generated at over-tidal(-harmonic), near-inertial and near-buoyancy frequencies.

*1.3. Sedimentary topography-slope perspective*

A challenge from geological perspective. According to Cacchione et al. (2002), in follow-up from Cacchione and Southard (1974), internal tides are the prime candidate for shaping the ocean's underwater topography, notably its average slope that has approximately the same value as the slope of internal tide characteristics. This is reasoned from the observation that, beyond continental shelves, the average seafloor-slope closely (critically) matches internal tidal characteristics slopes for mean $N \approx 2\times10^{-3}$ s$^{-1}$ around mid-latitudes. Cacchione et al. (2002), considering the upper 1000 m of the ocean mainly, assume N is constant at greater depths, which ignores the continued gradual decrease with increasing depth. However, as will be demonstrated in Section 3, ocean's volume-weighted mean N is three times smaller than the mean value above.

Cacchione et al. (2002) postulate that sediment erosion, and thus prevention of sediment deposition, occur at slopes where the semidiurnal internal tidal slope critically matches that of topography. The internal-wave model by Cacchione and Wunsch (1974) suggests that for such matching slopes the near-bottom flow is strongest. However, their 1D model is based on low vertical mode linear internal waves, adopting only bed shear-stress as means for (inhibition of) resuspension of sediment. Thereby, the effect of plunging breaking waves is not considered (For the effects on sedimentation resuspension due to better known surface wave breaking, see: Voulgaris and Collins, 2000), besides neglects of spring-neap variability, stratification variability and 3D-effects of topography.

As for seafloor topography, the advancement of observational techniques including multibeam acoustic echosounder and satellite altimetry have considerably improved mapping (e.g., Smith and Sandwell, 1997). Using such maps, a global ocean seafloor indexation has been compiled to an overall resolution of 1′ (1852 m in latitude) by Costello et al. (2010). An interesting finding of theirs concerns the separation of ocean area and volume per depth zone. (A correction to area and volume calculations is published in Costello et al. (2015), including a proper definition of depth zone). Whilst 11% of the ocean area and <1% of its volume is occupied by water depths <1000 m, the remarkable results are for the deep sea. It is found that 75% of the area and 90% of its volume are in the depth zone with water depths between $3000 < H < 6000$ m. A large part of this depth zone can be found in the abyssal hill's areas of the Pacific and Atlantic oceans. Only 4.4% of the ocean area and 1.9% of its volume are occupied by the depth zone with water depths between $1000 < H < 2000$ m. So, too little topography is in the depth zone of mean N to maintain ocean stratification following the reasoning around (6). Expanding to a depth zone of $100 < H < 2000$ m, the values are 10% and 2.3%, respectively. We recall that it is the sloping seafloor where most internal waves break and generate turbulence, not the ocean-interior.

**2. Materials and methods**

The foundation of topography-internal wave interaction leading to our ocean turbulence investigation has been an almost three decades-long observational program of a traveling mooring including instrument development and manufacturing. At some 25 sites (Fig. 1) distributed over the global ocean of varying topographic slopes one or more vertical mooring lines were deployed holding custom-made high-resolution low-noise temperature (T-)sensors (van Haren, 2018). The sites showed a large variety in seafloor topography, from abyssal plains to steep canyons, deep trenches, fracture zones, narrow ridges, large continental slopes, and seamounts. Here, sites shallower than continental shelves are not considered, and specific topics like internal wave interaction with sediment waves are not treated (for an example of such see, e.g., van Haren and Puig, 2017).

*2.1. Moored T-sensors*

Some NE-Atlantic sites like Mount Josephine, Rockall Trough and Faeroer-Shetland Channel were occupied with one or more moorings multiple times. The mooring lengths varied between 30 and 1130-m long strings holding a range of 30-760 stand-alone T-sensors at 0.5-2 m intervals, starting between 0.5 and 8 m from the seafloor. The duration of underwater deployment was at least five days, typically several months, and up to three years. The typical mooring was 100-150 m high with 100 T-sensors and was underwater for several months, sampling at a rate of once per second, resulting in the resolution of most energy-containing internal wave and turbulence scales. This allowed for calculation of turbulence values that were averages over most of the relevant scales, in the vertical and over at least inertial and tidal periods include spring-neap cycle. All moorings were held tautly upright after optimizing sufficient buoyancy and low-drag cables, for near-Eulerian measurements. All T-sensors were synchronized every 4 hours to a standard clock, so that vertical profiles were measured almost instantaneously within 0.02 s.

The main purpose of the moored T-sensors was to infer turbulence values using the sorting method of Thorpe (1977) over deep-sea topography under varying conditions of elevation, stratification, and local water-flow. The moored instrumentation and data processing are extensively described elsewhere (e.g., van Haren and Gostiaux, 2012; van Haren, 2018).

Near every mooring, one or more shipborne water column profiles were made using a Conductivity-Temperature-Depth (CTD) package, mainly SeaBird-911. The CTD-data are used to establish the local temperature-density relationship around the depth-range of moored T-sensors, and for reference of absolute temperature and large-scale stratification.

*2.2. Seafloor elevation data*

Topographic data are retrieved from external data depositories following pioneering works by Smith and Sandwell (1997). Such depositories are GEBCO (https://www.gebco.net) and EMODnet (https://emodnet.ec.europa.eu/en/bathymetry). These data are distributed to grids, e.g., GEBCO-2023 to 15″ (463 m in latitude, North-South direction), and are composites of data from satellite altimetry, shipborne single- and multi-beam acoustic echosounders, and from numerical estimates. Progress is made from the manual soundings of a century ago, via acoustic single-beam echosounder profiling tracks used by Bell (1975a,b), to present-day multibeam mapping and composite global mapping of 1′ (Costello et al., 2010) and smaller. Although the depositories rapidly fill with new high-resolution topographic data following modern multibeam surveys, less than 10% of the seafloor has been mapped at a resolution O(10-100) m so far. It will take at least several decades before the entire seafloor has been mapped at this resolution, if ever. It is noted that the 1′-resolution in some remotely sampled areas results from extrapolated data, while other areas are sampled at 10-100 times higher resolution.

Here, we investigate in some detail topography at two specific sites (Fig. 2) of NE-Atlantic Mount Josephine and E-Mediterranean West-Peloponnese using a variety of 15″-resolution GEBCO data, in conjunction with 3.75″ (116 m in latitude) resolution EMODnet data, and about 1.6″ (50 m in latitude) and 0.375″ (11 m in latitude) resolution multibeam data. The latter are obtained locally around the T-sensor mooring sites only. Our multibeam data are de-spiked and somewhat smoothed reducing the original sampling rate by a factor of two approximately. The scale variations allow for a limited investigation in slope dependence on horizontal scales.

Although the multibeam echosounder surveys were only a support-part of respective research cruises and therefore do not cover large areas, their extent is sufficient to resolve all internal wave scales. As a bonus, multibeam data processing also delivers information on the reflective properties of the substrate in the property of acoustic backscatter strength. This information was used by van Haren et al. (2015) to demonstrate that over Mount Josephine hard substrates consisting of coarse grain sizes and/or compacted sediment were almost exclusively found in areas with seafloor slopes that were supercritical for semidiurnal internal tides. Less reflective soft substrates consisting of fine grain sizes and/or water-rich sediment were found at sub-critical slopes.

The x-y 2D-gridded, being essentially 3D with z included, seafloor elevation data from above sources will be investigated spectrally, to compare with the 1D-single track, essentially 2D with z included, data from NE-Pacific hills explored by Bell (1975a,b). Some slope statistics is also pursued after computation of the proper slope at each 2D-gridded data-point to characterize the ratio (percentage) of slopes exceeding a particular value.

## 3. Results

The two small deep-sea areas for in-depth investigation are around the same mid-latitudes but otherwise distinctly different (Fig. 2). The NE-Atlantic is known for dominant semidiurnal internal tides besides an-order-of-magnitude smaller amplitude near-inertial waves. The East-Mediterranean lacks substantial tides so that internal waves are dominated at near-inertial frequencies only. In the NE-Atlantic, sampling sites are above the eastern side of large underwater Mount Josephine, about 400 km West of Southern Portugal. In the East-Mediterranean, the site is about 20 km West of Peloponnese, Greece.

*3.1. Vertical profiles*

The shipborne water column CTD observations demonstrate moderately stratified and seldom homogeneous waters in the lower 300 m above the seafloor when $N = N_{10}$ is computed over small-scale 10-m vertical scales (Fig. 3a). To get some idea of variability of layering in the ocean, $N = N_{100}$ 100-m scale profiles are given for the NE-Atlantic area in Fig. 3b, for comparison.

In the lower 250 m above seafloor of the East-Mediterranean site, the variability in N-profiles becomes larger than in the interior above, with average values slowly decreasing with depth (Fig. 3a). In the lower 100 m above the seafloor the variability in stratification is largest, both in the vertical and in time. The lower 400 m above the seafloor demonstrate >100-m averaged values $<N_{10}> = 7\pm6\times10^{-4}$ s$^{-1}$, with a gradual decrease of values from 8 to $6\times10^{-4}$ s$^{-1}$ towards the seafloor.

Similar mean N-values are observed above Mount Josephine, but near mean H = 3900 m where $<N_{100}> = 7\pm2\times10^{-4}$ s$^{-1}$ is observed in the lower 400 m above the seafloor (Fig. 3b). The observed errors and natural variations in N are about twice larger than sketched in Section 1.2. Thus, the NE-Atlantic site is about half-one order of magnitude more stratified, 2-3 times larger in buoyancy frequency, at a given pressure level compared to the East-Mediterranean, and may thus support more internal wave-energy and -shear.

At the deepest NE-Atlantic site considered here, albeit having the mean ocean water depth, the bottom slope is generally subcritical for internal tides (van Haren et al., 2015). This results from the gradual decrease of stratification with depth that leads to a steepening of internal wave characteristics following (1) while the seafloor slope generally becomes smaller for concave topography. To become (super)critical for semidiurnal internal tides, the local slope would need to be (larger than) about 10°.

The (semidiurnal tidal) supercritical portion found above Mount Josephine between 1000 < H < 2300 m (van Haren et al., 2015) may be part of the 5% of surface-area required to maintain the ocean stratification following (4). However, it is not sufficient alone, as this required 5% is larger than the global total of 4%-surface area of depth zone 1000 < H < 2000 m (Costello et al., 2010; 2015). Because also not all slopes between 1000 and 2000 m are expected to be supercritical (for semidiurnal internal tides even under $N > 2\times10^{-3}$ s$^{-1}$; Recall that basically all slopes are supercritical for particularly directed near-inertial waves), other supercritical slopes are sought. Supercritical slopes are more easily found at 100-2300 m relatively shallow depths, given the statistically larger N and thus smaller (1), (2), and sufficient turbulence by internal wave breaking.

How and what do internal waves shape the seafloor? Either the seafloor shape is concave formed by erosion mid-slope, which, given the general mean stratification profile, leads to less likelihood of turbulent mixing due to internal wave breaking in the deep, or it is convex by erosion above and below, which may favour deeper, supercritical slopes and associated enhanced stretches of mixing. Following nonlinear internal wave 2D-modelling, no distinct difference in turbulence intensity is found between internal wave breaking at convex, concave, or planar slopes (Legg and Adcroft, 2003). However, their model results show relatively large values of energy dissipation at sub-critical slopes, which are not found in ocean observations (e.g., van Haren et al., 2015). It is noted that the above modeling is based on 2D spatial-shapes and ocean topography is essentially 3D, like internal wave propagation and turbulence development. It is thus more generalizing to use full 3D seafloor elevation, i.e. full 2D-slope statistics, and evaluate internal wave breaking with that. While near continental margins, where the continental shelf dives into the continental slope around H = 200 m, the seafloor generally has a convex shape, it generally becomes concave at greater depths. Such topography would favour relatively shallow supercritical slopes (for semidiurnal internal waves).

*3.2. Seafloor spectra*

The multibeam echosounder data (Fig. 2) allow for local seafloor investigations, in particular on scale-size and -slopes, that go beyond those of Bell (1975a,b) who used single-beam echosounder data resolving height elevations at horizontal scales O(100) m, with a cutoff at Nyquist wavenumber of about $k_{Nyq}$ = 2.5 cpkm. Bell (1975a,b) established a general seafloor-elevation spectral fall-off rate of $k^n$, $-2 < n < -5/2$, for $k_0 < k < k_{Nyq}$, with a small $-1 < n < 0$ for $k = k_0 < 0.1$ cpkm and flattening $n = 0$ for $k < 0.01$ cpkm.

Our multibeam, EMODnet and GEBCO data-sets show a significantly steeper general fall-off rate in elevation spectra (Fig. 4a) than in Bell (1975a,b), with dominant low-wavenumber fall-off at a rate of $\sim k^{-3}$, and a saturation to noise values for $k > 10$ cpkm. The steeper spectral fall-off rates may be interpreted as a deviation from random distribution of seafloor elevation, in which energy is no longer uniformly distributed but favoured at the low-wavenumber side and reduced at the high wavenumber side. It has an intermittent appearance (Schuster, 1984).

In detail, the range between $\sim 0.3 < k < \sim 2$ cpkm shows the steepest fall-off rates $k^n$, $n < -3$, in the East-Mediterranean, and with extended steep slopes in the NE-Atlantic. The roll-off to weaker slopes for low wavenumbers is barely resolved, although the spectra do show the same tendency as in Bell (1975a,b). Extended GEBCO_2023-data across the Mid-Atlantic Ridge do resolve and show the roll-off at low wavenumbers (not shown).

The steepest fall-off rate is best visible after scaling the spectra, with $k^{-3}$ in Fig. 4b, to better indicate this slump-down in seafloor elevation. Noting that this slump-down does not indicate a spectral gap, the range of strong (steeper) deviation before resuming $n = -2.5$ or $-3$ centers around (tidal) transition wavenumber $k_T \approx 0.5$ cpkm, i.e. a wavelength of $L_T \approx 2$ km. It lies in the range of band-broadening that is visible in the data of Bell (1975a,b), and which overlaps with the range of internal wave generation in the abyssal NE-Pacific.

Here, it indicates a loss of seafloor topography variance at wavenumber $k > k_T$, by about one order of magnitude (measuring the wavenumber range between the two $n = -3$ spectral slopes). As the $k_T$ associates with that of the largest internal wave (excursion) scales, one may speculate that the loss of slope elevation relative to general fall-off rate is related with low-frequency internal waves, internal tides, mainly, or near-inertial waves, and their erosive turbulence generation smoothing out topography sizes. The spectral slump-down around $k_T$ is statistically significant. However, it also significantly varies between different sites, shifting about half an order of magnitude to higher wavenumbers at Mount Josephine. For general

GEBCO_2023 data from across the Mid-Atlantic Ridge in comparison with our data, one finds also strict k$^{-3}$ spectral fall-off rate and the spectral slump-down shifts by half an order of magnitude to lower wavenumbers (not shown).

*3.3. Slope statistics*

The mean slope of seafloor topography calculated by Bell(1975a,b) at >100-m scales is 3±1°. In our two small areas around the same latitude (Fig. 2), the slope distribution varies with length-scale; at slopes > 3°, the shorter the length-scales the steeper slopes are calculated (Fig. 5). The 3°-slope is the median seafloor-slope value for our small East-Mediterranean site, irrespective of scales used (Fig. 5a). The median seafloor-slope value for our small NE-Atlantic site varies per scale and is about 4° at 1′-resolution and 5° at 0.25′-resolution (Fig. 5b). These slopes are found for common mean N at z < -2000 m.

For most of the global ocean using a 1′-resolution, Costello et al. (2010) find that 9.4% of the seafloors have a slope between 2 and 4°, 8.2% have a slope > 4° and 4.5% have a slope > 6°. From which we conclude that 3.7% have a slope between 4° and 6°. Recall that 75% of the ocean area and 90% of its volume has seafloors between 3000 and 6000 m, where thus most of these slope(percentage)s are.

At our two sites, slopes are slightly steeper at 1′-resolution, and 5% of the seafloors have slopes > 10°. For the East-Mediterranean, at 1′-resolution the (semidiurnal tidal) critical slope is 9.6° for local N = 6.6×10$^{-4}$ s$^{-1}$. This value of N is to within error the same value found after volume-weighted averaging below z < -500 m of near-surface seasonal stratification variation for open NE-Atlantic and Mariana Trench (Pacific Ocean) CTD observations (Fig. 6). It is noted that this averaging includes effects of Antarctic Bottom Water, observed in the Mariana Trench profile.

At this volume-weighted average N, associated supercritical slopes for semidiurnal lunar internal tide are found to occur for 5% of all slope values, i.e. 5% of all slopes are > 9.6° (Fig. 5a). This percentage is reached at an angle of 20° for 1.6″-resolution multibeam data of Mount Josephine (Fig. 5b). As this angle-value is steeper than that for sedimentary stability, its seafloor texture may thus foremost consist of hard substrate, which indeed has been observed in multibeam data for supercritical slopes (van Haren et al., 2015).

However, when computed at the larger 1′-resolution (magenta graphs in Fig. 5) one finds a 5%-transition for 11°, which is to within error the same slope for NE-Atlantic and East-Mediterranean multibeam data. Thus, seafloor slope statistics are identical to within error between our two sites at 1′-resolution. Recall that the 1′-resolution is close to the spectral transition length-scale $1/k_T$ (Fig. 4), and close to the internal tidal (largest internal wave) excursion length. The characteristics (2) slope-range between 9 and 11° comprises the 10° seafloor-slope above which semidiurnal internal tides become supercritical at sites of the 3900-m mean ocean water depth given local N. Although this N is found around 1100 m in the East-Mediterranean, the approximate 2.5 times weaker N, compared with NE-Atlantic data at the same depth, associate with the lack of tides so that inertial energy is about 40% of total (inertial and tidal) internal wave energy found in the NE-Atlantic.

Although volume-weighted mean N and thus mean internal tidal characteristics slopes are found in the depth zone of mean ocean depth comprising 75% of the ocean area (Costello et al., 2010), the occurrence of 5% supercritical slopes diminishes local relatively weak turbulence dissipation rates $O(10^{-9})$ $m^2s^{-3}$ to a small contribution <10% to maintain global ocean stratification. However, the coincidence of 5% seafloor slopes with supercritical slopes for volume-weighted mean N and N at mean ocean depth may reflect an ocean-wide balance of internal wave-turbulence and topography interaction. Locally in the deep-sea, such turbulence may have considerable influence for redistribution of sediment and nutrients. Examples are short-term contributions of inertial waves generated by, e.g., large storms such as typhoons. For standard mean $10^{-8}$ $m^2s^{-3}$ over h = 200 m in H = 3900 m, 5% semidiurnal tidal supercritical slopes yield a global contribution of $2.5\times10^{-11}$ $m^2s^{-3}$. A passing typhoon may dissipate $10^{-7}$ $m^2s^{-3}$ as observed (van Haren et al., 2020) over h = 200 m in H = 3100 m at any slope, as basically all slopes are super-critical for a near-inertial wave characteristic.

Considering the more numerous moored T-sensor observations from the depth zone 100-2000 m, which comprises 10% of the ocean area (Costello et al., 2010; 2015), the larger local N yields 50% of local slopes to be supercritical (for semidiurnal tides), cf. Fig. 5. The overall 50% of 10% = 5% supercritical slopes are sufficient to maintain the entire ocean stratification over typical h = 100 m in average H =1000 m and observed turbulence dissipation rate (6). This is also the depth zone in which cold water corals (CWC) thrive (United Nations, 2017). The suspension feeding CWC rely on nutrient supply via sufficiently turbulent hydrodynamic processes.

**4. Discussion**

Cacchione et al. (2002) made calculations with water-level height mean N above a mean-depth ocean seafloor, holding mean-N value constant for all z < -150 m. This is not representing a realistic internal wave turbulence environment, because (1) N monotonically decreases with depth (except under local conditions such as when Antarctic Bottom waters are found near 4000 m), (2) internal waves refract so that local N has to be accounted for, (3) internal waves predominantly break at sloping seafloor and not in the ocean-interior.

Deep-ocean internal waves can be modelled to first order as linear waves and are found ubiquitous throughout all seas and oceans. However, given their natural environment which is not constant in space and time and their potential interactions with other water-flows they divert considerably from linear, constant-frequency waves. First, the stratification-support varies under internal wave straining, boundary flows and (sub-)mesoscale eddies, so that N shows a relative variation of typically ±20%. Although semidiurnal internal tides are dominant energetically, their variation in frequency alone provides 6% variations in slope of characteristics. In the deep-sea, roughly the deeper half of all oceans, N < 8f, at mid-latitudes, and full internal wave equations show a spread in internal tide characteristics of >15%. All these natural variations, not counting variations in seafloor slope determination as a function of length-scale, provide a relative error in dominant internal wave characteristics of 25-50%.

It thus seems impossible to find a particular persistent critical slope for a given single internal wave frequency on a 1′-scale, and which also ignores the highly nonlinear character of dominant turbulence-generating upslope propagating bores, which are composed of motions at many internal wave frequencies. Therefore, it is not surprising that most internal wave generated turbulence occurs at (just) super-critical slopes, which provides a broader slope- and thus frequency-range.

As the above relative uncertainty range of internal wave characteristics matches that of relative error of about 33% in mean turbulence dissipation rates, it reflects the uncertainty in determining present-day percentage of supercritical slopes required to maintain ocean stratification, being 5±1.5% for seas where internal tides dominate. This uncertainty also sets the bounds for robustness of the internal wave-topography interactions: It is the margin within which variations are expected to find sufficient feed-back not to disrupt the system from some equilibrium.

If so, spiking any variations to this system must go beyond an energy variation of about 30%, which is larger than (the determination of) tidal variation but probably less than inertial motions variations, say wind (Wunsch and Ferrari, 2004). In terms of stratification, 30% of variation is feasible near the sea-surface via seasonal but also day-night variations, but is well exceeding any natural stratification variations in the deep-sea, say for $z < -500$ m.

Suppose we can go beyond 30% variation, what will happen then? It takes at least decades-centuries-millennia for the seafloor elevation to adapt to an equilibrium of sufficiently supercritical slopes. If N increases by >30% uniformly throughout the ocean, more (higher frequency) internal waves will be supported and internal tide characteristics will become flatter. As a result, more (unaltered) seafloor slopes will become supercritical, which will raise the amount of ocean turbulence. Perhaps by >30%. Increased turbulence means more heat transport, hence a reduction of N that diminishes the initial 30%-increase. It goes without saying that the opposite occurs in the event of a decrease in N.

Although being a meagre proof of evidence, we recall that the (East-)Mediterranean deep-sea has a factor of 2-3 times weaker N than the tidally dominated NE-Atlantic Ocean at any given depth. This factor is commensurate with the 2-3 times weaker near-inertial internal wave energy compared with that of combined energy of internal tide and near-inertial waves. Both sea-areas are in present-day equilibrium. As a result, it seems that it is not the buoyancy (density stratification) variations that strongly disturb the equilibrium, but the external sources of internal wave (kinetic energy).

(Just-)super-critical slopes are probably bounded by a maximum of 15° for sedimentary slope-instability. At any rate, the super-critical seafloor slopes allow development of upslope propagating bores and rapid restratification of the back-and-forth sloshing internal tides. In contrast with foreward wave-breaking at a beach, internal tides break backwards at a slope (van Haren and Gostiaux, 2012). This may explain a lack of clear swash, i.e., while upslope propagating bores may be considered as uprush, a clear vigorous backwash is not observed during the downslope warming tidal phase near the seafloor in moored T-sensor data. This demonstrates a discrepancy with the 2-D modelling of Winters (2015). While in the model most intense turbulence is found near the seafloor during the downslope phase expelling into the interior, ocean observations demonstrate largest turbulence around the upslope propagating backwards breaking bore, with the bore sweeping material up from the seafloor (Hosegood et al., 2004). Probably some 3D- or rotational aspect is important for ocean internal wave breaking, yet to be modelled.

We have considered a combination of seafloor elevation and internal wave turbulence data to revisit the interaction between topography and water flows in the deep-sea. From various perspectives including turbulence values, vertical density stratification, (water) depth zones, and seafloor and internal wave slopes and scale-lengths it is found that interaction is relatively stable, whereby mainly internal tides and near-inertial waves shape the topography to within 30% variability.

**5. Conclusions**

The median value of seafloor slope of 3±0.2° from multibeam echosounder and satellite data from NE-Atlantic, mid-Atlantic Ridge and East-Mediterranean sites closely matches the half-a-century-ago established rms-mean slope of 3±1° from single-beam echosounder data across PE-Pacific abyssal hills (Bell, 1975a,b). Our result is found only weakly dependent of scale, which we varied between 0.027′ and 1′. It lends some robustness to the determination of seafloor slopes, for horizontal scales that match internal wave excursion lengths.

The average spectral fall-off rate of seafloor elevation is found steeper than Bell's (1975a,b), which indicates a non-uniform distribution of scales instead of a uniform distribution as previously suggested. In particular, the spectral slump-down around a length-scale of 2 km is noted, which suggests a lack of seafloor elevation shaped by the largest internal wave excursion length dissipating its energy into turbulence creating sediment erosion. This spectral slump-down is found in seafloor elevation data from both the NE-Atlantic, where semidiurnal internal tides prevail, and from the East-Mediterranean, where tides are small and near-inertial motions dominate internal waves. In the East-Mediterranean, the buoyancy frequency is found smaller than at corresponding depths in the NE-Atlantic, which is commensurate with the contribution of internal tides (and lack thereof).

Recent moored high-resolution T-sensor data demonstrated that internal wave breaking is found most vigorously above seafloor slopes that are supercritical rather than much more limiting critical for (semidiurnal) internal tides, with local turbulence dissipation rate $> 10^{-7}$ $m^2s^{-3}$ in depth zone $100 < H < 2200$ m (NE-Atlantic). This depth zone hosts most of cold-water corals that depend on vigorous turbulence for nutrient supply. Our seafloor statistics show that 50% of the slopes are supercritical for stratification in this depth zone, which compensates for the depth zone's occupation of only 10% of ocean area. As a result, we find that internal wave breaking at 5±1.5% of slopes suffices to maintain global ocean density stratification.

In greater depth zones, internal wave breaking is generally less turbulent contributing <10% to maintain global stratification mainly due to steeper internal tidal characteristics slopes. Even in the deep-sea however, 5% of seafloor slopes coincide with supercritical slopes for volume-weighted mean N, and N at mean ocean depth, supporting an ocean-wide balance of topography and internal wave-turbulence interaction. Turbulence may be locally important for the redistribution of heat, nutrients, and oxygen, e.g., during the passage of typhoons generating near-inertial waves as most seafloor slopes are supercritical for (one characteristic of) such waves, also in great depth zones.

**Ackowledgments**

NIOZ T-sensors were supported in part by NWO, the Netherlands Organization for the advancement of science. We thank NIOZ-MRF and the captains and crews of the R/V's Pelagia and Meteor, as well as from numerous other research vessels we joined, for their very helpful assistance during mooring construction and deployment.

**Declaration of competing interest**

The authors have no conflicts to disclose.

**Data availability**

The data that support the findings of this study are available from the authors upon reasonable request. Seafloor elevation data are extracted from depositories https://www.gebco.net and https://emodnet.ec.europa.eu/en/bathymetry. Raw East-Mediterranean CTD data supporting the results of this study are available in database https://data.mendeley.com/datasets/6td5dxf6bj/1.

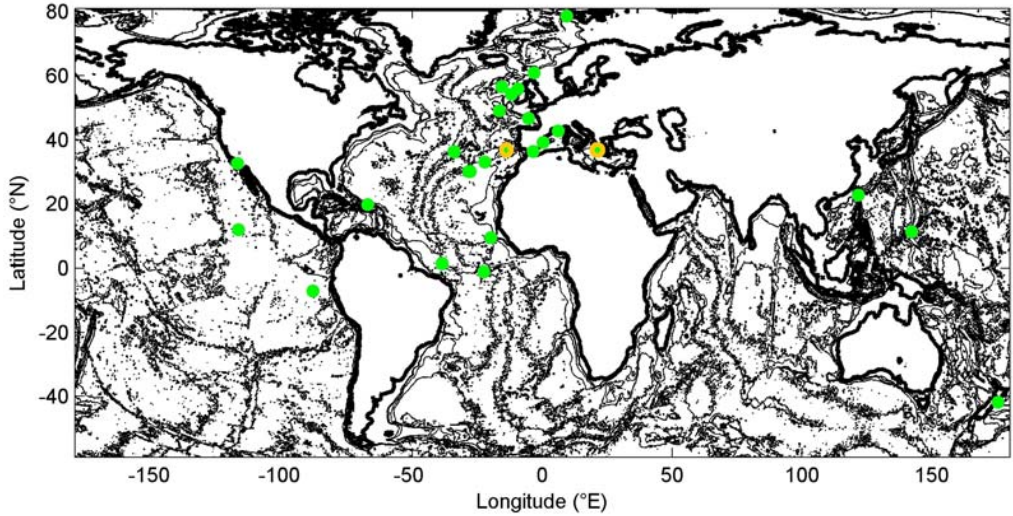

**Figure 1**. Global map of seafloor topography (1′-version of Smith and Sandwell, 1997) with contours every 1500 m together with sites (green dots) of NIOZ T-sensor moorings for deep-sea turbulence and internal wave research. The two orange encircled sites are discussed in some detail in the present paper.

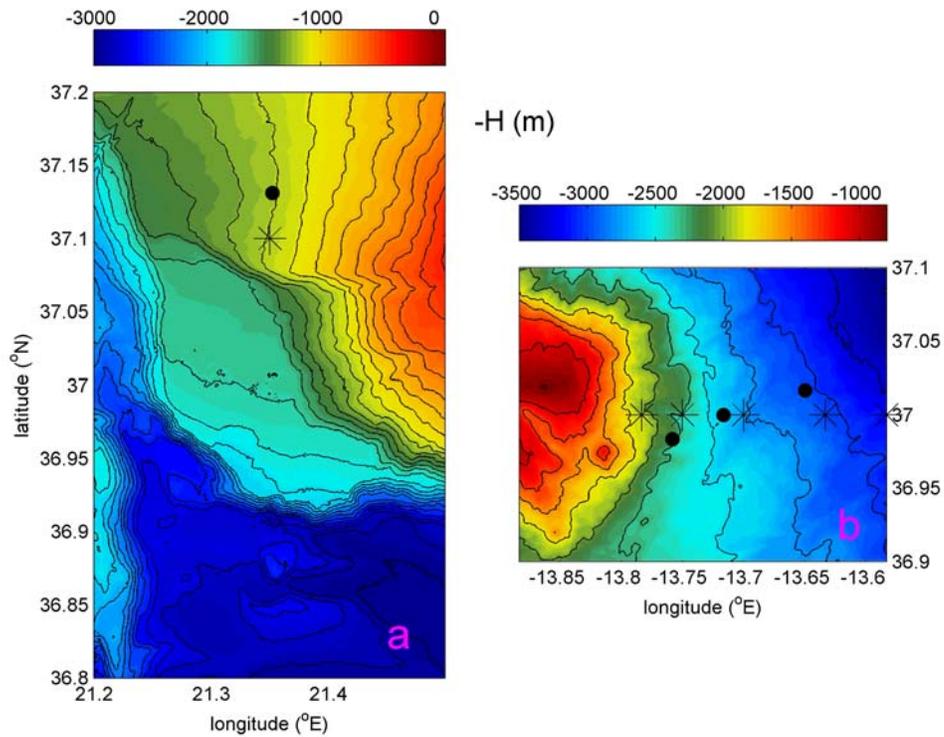

**Figure 2**. Two detailed maps made by shipborne multibeam echosounder. (a) Part of West-Peloponnese continental slope, Greece, East-Mediterranean Sea, from R/V Meteor. Black dot is moored T-sensor location, star indicates yoyo-CTD station. (b) Part of eastern slope of Mount Josephine, Northeast-Atlantic Ocean, from R/V Pelagia. Black dots indicate moored T-sensor locations, stars indicate CTD stations during various years.

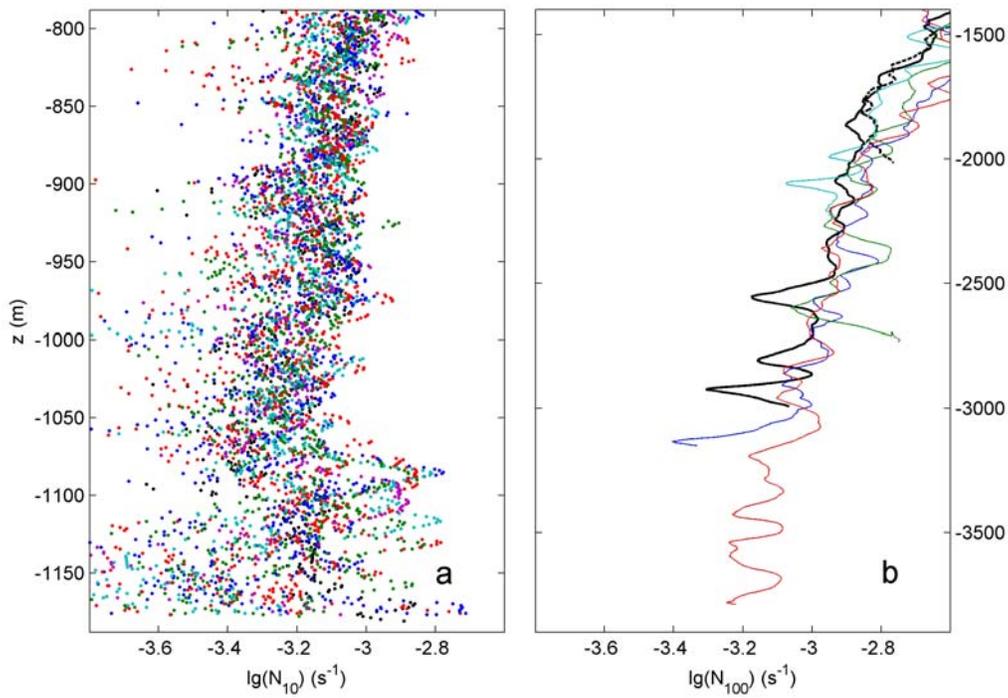

**Figure 3**. Logarithm of buoyancy frequency from shipborne water column CTD. (a) One inertial period of 400-m high yoyo-CTD hourly profiles, 19 in total. Computations are made over 10-m vertical scales from fixed location at H = 1180 m water depth in the East-Mediterranean. (b) Vertical range of 2500 m of profiles. Computations are made over 100-m vertical scales down to 10 m from various seafloor depths of the eastern slope of Mount Josephine, NE-Atlantic. X-axis scale is identical, but Y-axis scale is different compared to a.

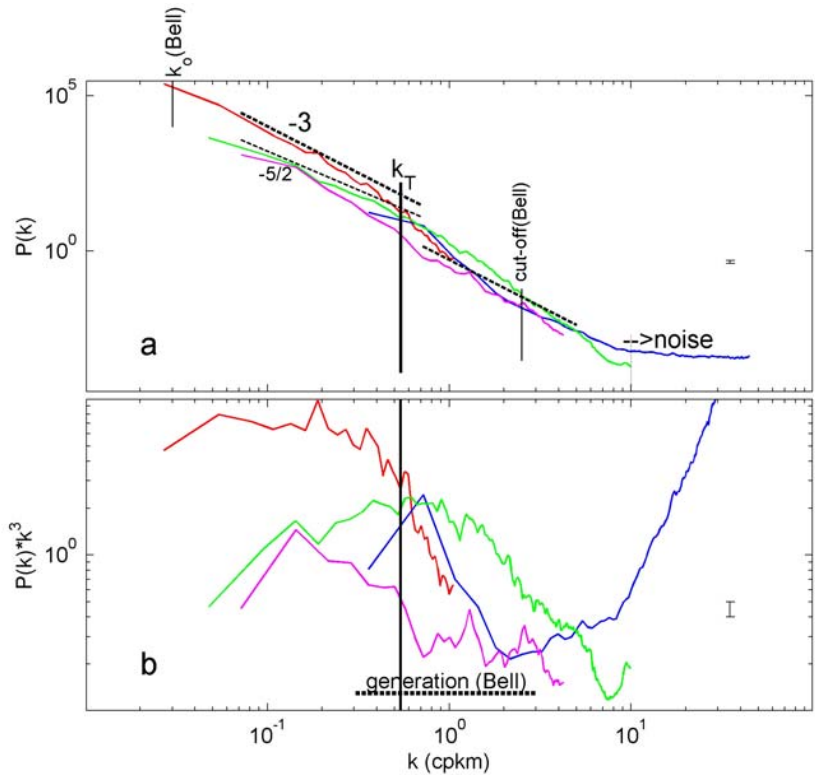

**Figure 4**. Spectral analysis of seafloor elevation as a function of horizontal wavenumber k (the inverse of horizontal length-scale L). (a) On the log-log plot matching $k^{-3}$ (-3 on log-log plot) spectral slopes are represented by straight lines, with $k^{-5/2}$ (-5/2-slope) the slope reported by Bell (1975a). Bell's low-wavenumber cut-off is indicated, albeit barely resolved, as well as high wavenumber roll-off $k_0$. The central vertical line at transient wavenumber $k_T$ indicates a wavelength of 1852 m (1′ in latitude). The green plot is for 1.6″ sampled Mount Josephine (NE-Atlantic) multibeam data, the three other spectra are for the East-Mediterranean: 0.375″ sampled multibeam data (blue), 3.75″ sampled EMODnet data (magenta), and 15″ sampled GEBCO data (red). (b) The same as a., but spectra scaled with $k^3$, the dominant low wavenumber slope. Bell (1975a,b)'s one-decade range of internal wave generation of Pacific abyssal hills is indicated.

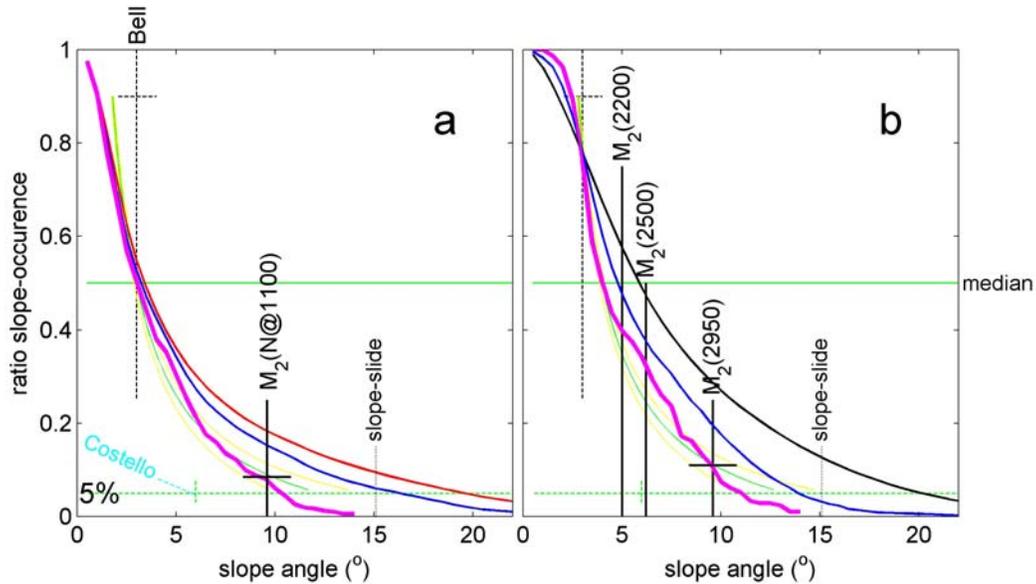

**Figure 5**. Seafloor slope statistics curves of ratio of angle-excess occurrence, for different seafloor-elevation sampling scales. For comparison, normalized curves are plotted for internal wave characteristic slopes in green using (1) and in yellow using (2). The horizontal center green line indicates the median, the lower dashed-green line indicates the 0.05(5%) level that is required for (just-)supercritical slopes to generate sufficient global turbulence. Vertical black-dashed lines indicate Bell (1975b)'s average slope of Pacific abyssal hills and the maximum slope before collapse of sediment packing 'slope-slide'. (a) From East-Mediterranean map of Fig. 2a using three different scales for elevation slope-computations: L = 1′ (magenta), 15″ (blue) and 3.75″ (red). Costello et al.'s (2010) range of percentages is indicated in light-blue (see text). The semidiurnal lunar ($M_2$) internal tide slope is indicated for local N at H = 1100 m, with corresponding error/spread in characteristics value. (b) From NE-Atlantic map of Fig. 2b using L = 1′ (magenta), 14.5″ (blue) and 1.6″ (black). $M_2$-slopes for three different mooring sites are indicated.

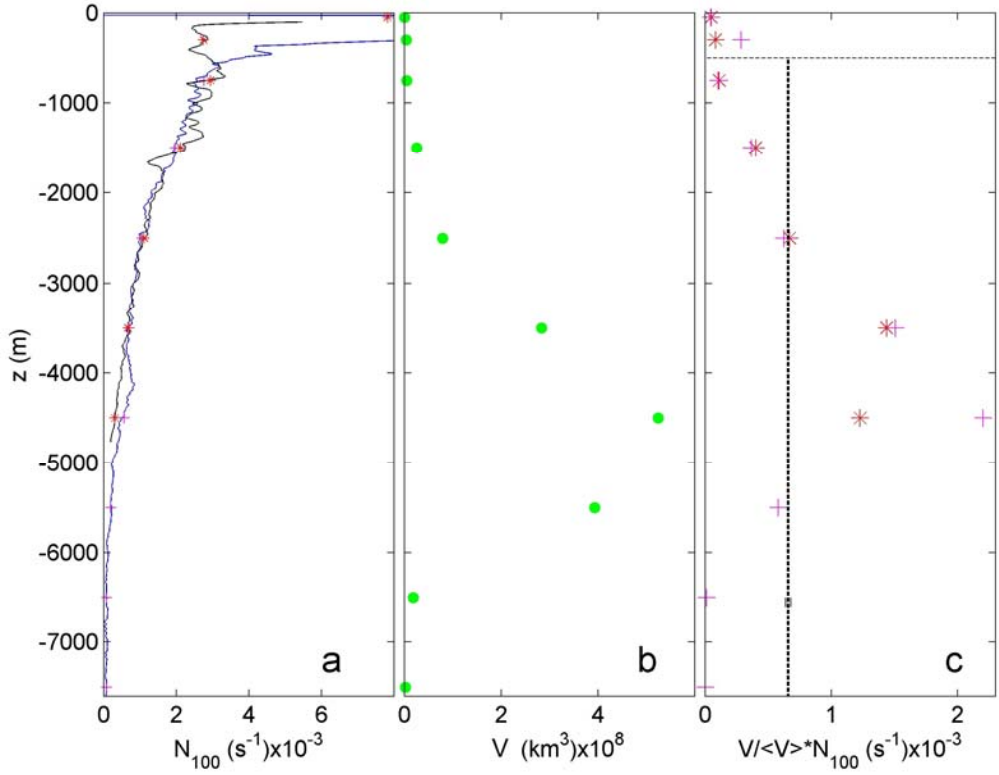

**Figure 6**. Weighted function of buoyancy frequency as a function of the vertical. (a) Large-100-m scale buoyancy frequency from NE-Atlantic, off Mount Josephine (green profile with red stars) and over and inside Mariana Trench (blue with magenta plusses). (b) Ocean volume (V) per depth zone, adapted from Costello et al. (2010) with depth zones defined in Costello et al. (2015). (c) Volume-weighted values of buoyancy frequency in (a) using (b). $\diamond$ indicates averaging. Note the relatively large value from around -4500 m in Mariana Trench data that is associated with Antarctic Bottom Water. Below $z < -500$ m, the vertical dashed line indicates mean value $V/\langle V \rangle \cdot N_{100} = 6.6 \pm 0.2 \times 10^{-4}$ s$^{-1}$, for both profiles.